\documentstyle[buckow]{article}
\def\a{\alpha} 
\def\b{\beta} 
\def\g{\gamma}  
\def\d{\delta}  
\def\e{\epsilon} 
\def\h{\eta}
\def\l{\lambda}

\def\m{\mu}  
\def\n{\nu}  
\def\r{\rho}
\def\f{\phi}

\def\ve{\varepsilon} 
\def\ca{{\cal A}}

\def\co{{\cal O}}

\def\pa{\partial}
 
\def\ll{\left} 
\def\rr{\right} 
\def\be{\begin{equation}}  
\def\ee{\end{equation}}  
\def\beq{\begin{eqnarray}}  
\def\eeq{\end{eqnarray}}  
\def\nn{\nonumber}
\newcommand{\bqn}{\begin{eqnarray}}\newcommand{\eqn}{\end{eqnarray}}
\newtheorem{theorem}{Theorem}[subsection]

\begin {document}
\begin{raggedleft}
ULB-TH-01/36 \\ [.5cm]
\end{raggedleft}

\def\titleline{
Multi-graviton theories : yes-go and no-go results \footnote{  
 Talk given at the RTN meeting 
``The Quantum Structure of Spacetime and the Geometric Nature of 
Fundamental Interactions'', Corfu, 13-20 September 2001.  }
}

\def\authors{
Nicolas Boulanger~\footnote{``Chercheur F.R.I.A., Belgium''}}

\def\addresses{
Physique Th\'eorique et Math\'ematique,  Universit\'e Libre  
de Bruxelles,  C.P. 231, B-1050, Bruxelles, Belgium  \\   
{\tt nboulang@ulb.ac.be}
}

\def\abstracttext{

We report yes-go and no-go results on consistent cross-couplings for
a collection of gravitons.
Motivated by the search of theories where multiplets of massless spin-two 
fields cross-interact,
we look for all the consistent deformations of a positive sum of Pauli-Fierz
actions. We also investigate the problem of deforming a 
(positive and negative) sum of linearized Weyl gravity actions
and show explicitly that there exists multi-Weyl-graviton theories.
As the single-graviton Weyl theory, these theories do not have an energy 
bounded from below. 
 }

\large

\makefront

\section{Introduction}
In this report based on the works 
\cite{BDGH,Boulanger:2000ni,Boulanger:2001he} done in collaboration with
T. Damour, L. Gualtieri and M. Henneaux,
 we review some results about 
classical field theories involving several gravitons in interaction.
To be more precise, in the following we use the word ``graviton'' as referring
to a tensor $h_{\mu\nu}$ with two symmetric covariant indices. In the case of
Einstein gravity, the term is justified, while in the case of Weyl gravity 
it is no more exact, and the term ``Weyl-graviton'' will be of use.
The main question is to see whether it is possible to have a spin-two 
analogous of Yang-Mills theories, in the sense of multiplets of gravitons 
non-linearly
cross-interacting through some vertices in the Lagrangian.
In the context of Yang-Mills theories, the different massless spin-one fields
belong to the adjoint representation of the semi-simple gauge (Lie) algebra
and see each other through the Yang-Mills cubic vertex
(at first order in the coupling constant)
 ${\cal{L}}^{cubic}
\propto f_{abc}F^{\mu\nu a}A_{\mu}^bA_{\nu}^c$, where the 
``cubic coupling constant'' $f_{abc}$ are the structure constants of the group.
We may wonder if there exists similar theories involving multiplets of 
gravitons, each of these carrying an extra index running over some algebra-like
structure and where the various gravitons of the multiplet could feel each 
other through some vertices in the Lagrangian.

The idea for the construction of such multi-graviton theories is to start from
 a free theory given by the sum of $N$ separate, completely decoupled free
actions, each of these describing a single free graviton, and then try to 
perturbatively deform this free theory in order that the various gravitons
not only self-interact, but also cross-interact.
This is to say, we will look for all the possible interactions vertices to add 
to the free action, order by order in a deformation parameter, 
until a complete interacting theory is found.
We require that in the limit where the deformation parameter goes to zero,
the interacting multi-graviton theory (if any) smoothly goes back to the 
starting one.

Such an exhaustivity of the process (all the possible vertices!)
is made possible by using cohomology-BRST 
tools \cite{Barn-Henn}, in the Batalin-Vilkovisky formalism 
\cite{list}. On the one hand the Noether method is quite useful to build
an interacting gauge theory starting from a free one, but on the other hand
it is not systematic enough to allow the exhaustivity of the 
deformation : unless
you are very clever, you may miss some deformation vertices, and moreover
you are not sure you found them all. The BRST-based deformation method
\cite{Barn-Henn} relies on strong mathematical results, clearly organizes 
the calculation of the non-trivial consistent couplings in terms of 
cohomologies
already known or easily computed, and thus enables you to find all the possible
interactions vertices, in an exhaustive way.

When we say that a deformation is consistent, it means that the deformed 
interacting action possesses the same number of physical degrees of freedom, 
does not contain any inconsistency between the equations of motion, is free 
from ghosts, local, etc. In other words, if the starting theory was defined 
to be consistent, then so will be the deformed one. 


The plan is as follows : in section \ref{sect1} (based 
on \cite{BDGH,Boulanger:2000ni}) we give no-go and yes-go results about 
the consistent deformations of a positive sum of Pauli-Fierz actions. 
In section \ref{sect2} (based on 
 \cite{Boulanger:2001he}) we give no-go and yes-go 
results about the consistent deformations of a 
positive and non-positive sum of linearized Weyl actions and finally 
in section \ref{sect3} (based on the forthcoming work \cite{BHN}) we 
briefly comment
about multi-Weyl-graviton theories as gauge theories of non semi-simple 
extensions of the (super)conformal group. 
\section{Deformation of a sum of Pauli-Fierz actions}
\label{sect1}

It was shown by Pauli and Fierz \cite{Fierz:1939ix} 
that there is a unique, consistent
action describing a pure spin-2
massless field. This action happens to be the linearized 
Einstein-Hilbert action. Therefore, 
the action for a collection $\{h_{\mu \nu}^a\}$ 
of $N$ {\it non-interacting}, massless spin-2 fields in spacetime dimension
$n$ ($a = 1, \cdots, N$, $\mu, \nu = 0, \cdots, n - 1$) 
must be (equivalent to) the sum of $N$ separate Pauli-Fierz actions, 
namely\footnote{We use
the signature ``mostly plus'': $- + + + \cdots$.
Furthermore, spacetime
indices are raised and lowered with the flat Minkowskian metric
$\eta_{\mu \nu}$. Finally, we take the spacetime dimension $n$ to be
strictly greater than $2$ since otherwise, the Lagrangian is
a total derivative. Gravity in two dimensions 
needs a separate treatment.} 
\beq 
\label{startingpoint} 
S_0[h_{\mu \nu}^a] &=& \sum_{a = 1}^N \int d^n x \ll[ 
-\frac{1}{2}\ll(\pa_{\m}{h^a}_{\n\r}\rr)\ll(\pa^{\m}{h^a}^{\n\r}\rr) 
+\ll(\pa_{\m}{h^a}^{\m}_{~\n}\rr)\ll(\pa_{\r}{h^a}^{\r\n}\rr)\rr.\nn\\ 
&&\ll.-\ll(\pa_{\n}{h^a}^{\m}_{~\m}\rr)\ll(\pa_{\r}{h^a}^{\r\n}\rr) 
+\frac{1}{2}\ll(\pa_{\m}{h^a}^{\n}_{~\n}\rr) 
\ll(\pa^{\m}{h^a}^{\r}_{~\r}\rr)\rr] \, , \; \; n>2. 
\eeq 
Our treatment, which is purely algebraic, 
extends (at least formally) to the case where the the collection
$\{h_{\mu \nu}^a\}$ is, possibly uncountably, infinite (see \cite{BDGH}). 

The action (\ref{startingpoint})
 is invariant under the following linear gauge transformations, 
\be
\delta_\epsilon h^a_{\mu \nu} = \partial_\mu \epsilon_\nu^a 
+ \partial_\nu \epsilon_\mu^a 
\label{freegauge} 
\ee 
where the $\epsilon_\nu^a$ are $n \times N$ arbitrary, independent functions.
These transformations are abelian and irreducible.  
We rewrite the free action (\ref{startingpoint}) as 
\beq 
\label{PFalgebradelta} 
S_0&=&\int d^nx~k_{ab}\ll[ 
-\frac{1}{2}\ll(\pa_{\m}{h^a}_{\n\r}\rr)\ll(\pa^{\m}h^{b\n\r }\rr) 
+\ll(\pa_{\m}h^{a \m}_{~\n}\rr)\ll(\pa_{\r}h^{b \r\n}\rr)\rr.\nn\\ 
&&\ll.-\ll(\pa_{\n}h^{a \m}_{~\m}\rr)\ll(\pa_{\r}h^{b \r\n}\rr) 
+\frac{1}{2}\ll(\pa_{\m}h^{a \n}_{~\n}\rr) 
\ll(\pa^{\m}h^{b \r}_{~\r}\rr)\rr]\,, 
\eeq
with a quadratic form $ k_{a b}$ defined by the kinetic terms. In the 
way of writing the Pauli-Fierz free limit above, equation 
(\ref{startingpoint}),
$k_{ab}$ was simply the Kronecker delta $\d_{ab}$. What is
essential for the physical consistency of the theory (absence of
negative-energy excitations, or stability of the Minkowski vacuum) is that
$ k_{a b}$ defines a positive-definite metric 
in internal space; it can then be normalized to be  $\delta_{ab}$ by
a simple linear field redefinition. 

\subsection{No-go result}
Here comes the no-go result, with the assumptions clearly stated. 
\begin{theorem}
\label{t1}
Under the assumptions of : locality, Poincar\'e invariance, 
Eq.(\ref{startingpoint}) as free field limit
and at most two derivatives in the Lagrangian, the only consistent 
deformation of Eq.(\ref{startingpoint}) 
involving a collection of spin-2 fields is 
(modulo field redefinitions) a sum of
independent Einstein-Hilbert (or possibly Pauli-Fierz) actions, one
for each fields.
\end{theorem}
It means that no cross interactions are possible : no Yang-Mills-like
massless spin-2 theory.

\subsection{Yes-go result}

Relaxing the hypothesis on the number of derivatives in the Lagrangian, 
allowing for PT breaking terms ({\it{i.e.}} explicit presence of the 
completely antisymmetric Levi-Civita tensor) and restricting to $n=3$ 
spacetime, we can evade the previous no-go theorem and the positive sum
(\ref{PFalgebradelta}) admits a possible consistent deformation.
Schematically, the action (\ref{PFalgebradelta}) writes
$ S_0^{P.F.}[ h_{\m\n}^a ] \sim \int d^3 x \d_{ab} \pa h^a \pa h^b$
and is deformed as follows, at first order in the deformation parameter 
$\lambda$ (for an explicit writing of the deformed theory, 
see \cite{Boulanger:2000ni}):
\be
S_0^{P.F.}[h_{\m\n}^a]\rightarrow S_0^{P.F.}[h_{\m\n}^a]
+\lambda \int d^3x a_0 + \co (\l^2) 
\ee 
where $a_0$ is the cubic vertex which looks like
$a_0 \sim \ve^{\m\n\r}a_{abc}(\pa h^a \pa h^b \pa h^c)_{\m\n\r}$.
The constants $a_{abc}\equiv k_{ad}a^d_{~bc}$ play the same role as the 
structure constants in 
Yang-Mills theory, except that here it doesn't define a Lie algebra, but a
commutative, symmetric algebra $\ca$.
The fields $h_{\m\n}^a$ live in $\ca$ \footnote{the product between two 
elements $u,v$ of $\ca$ writes 
 $z^a=(u\star v)^a=a^a_{~bc}u^bv^c$, in a basis $\{ e_a\}_{a=1}^N$ of $\ca$.
In this basis, the commutativity and symmetry properties
read $a^a_{~bc}=a^a_{~cb}, a_{abc}=a_{(abc)}$, respectively 
}.
The gauge transformations are deformed, at first order in $\l$, to become
schematically 
\be
\d_{\e}h_{\m\n}^a=2\pa_{(\m}\e_{\n)}^a+\l[\ve^{\m\n\r}
(\pa h^a\pa\e^c)_{\m\n\r}a^a_{~bc}].
\ee
The point is that this deformation can be continued to all orders in $\l$,
giving rise to a complete exotic theory of non-linearly interacting multiplets
of $h_{\m\n}^a$-fields : 
\be
S^{EX}[h_{\m\n}^a]=S_0^{P.F.}[h_{\m\n}^a]+\l\int d^3x a_0+ \l^2 ...
\ee
The writing of the complete exotic action is done 
(see \cite{Boulanger:2000ni}) by using the first order
formulation of gravity, {\it{\`a la}} Chern-Simons 
\cite{Witten:1988hc&Deser:1981wh}


In the proof of theorem \ref{t1}, the algebra was constrained to be also
associative\footnote{in terms of the $a^a_{~bc}$, 
the associativity property reads $a^a_{~b[c}a^b_{~d]e}=0$}
. The positive-definiteness of the internal metric $k_{ab}$ 
 together with the commutativity, associativity and symmetry of $\ca$
implies that $\ca$ is a direct sum of one-dimensional ideals, implying 
in turn the 
existence of a basis where $a^a_{~bc}=0$ whenever two indices are different
\cite{BDGH} : the cross-interactions between the various gravitons can be
removed by redefinitions. This is why theorem \ref{t1} holds.
If $k_{ab}$ is of mixed signature, however,
the algebra $\cal A$ need not be trivial, and
one can construct truly interacting multi-gravitons theories \cite{Cutler:dv}.
Also, keeping $k_{ab}$ positive-definite but with $\ca$ no more associative, 
which is the case of the exotic theory, the interactions between the various
gravitons do exist. 

\section{Deformation of a sum of linearized Weyl actions}
\label{sect2}

The free theory for one Weyl-graviton is now 
\be
S_0[h^a_{\m\n}]=\frac{1}{2}\int d^4x {\cal W}_{\a\b\g\d} 
{\cal W}^{\a\b\g\d},
\label{freeactionconf}
\ee
for $g_{\m\n}=\h_{\m\n}+ \l h_{\m\n}$, $h_{\m\n}$ and $\l$ 
being dimensionless.
Here, $ {\cal W}^{\a}_{~\b\g\d}$ is the linearized Weyl tensor
constructed out of $h_{\m \n}$.
The free action (\ref{freeactionconf}) is invariant under both linearized 
diffeomorphisms and the linearized version of the Weyl rescalings,
\be
\d_{\h,\f}h_{\m\n}=\pa_{\m}\h_{\n}+\pa_{\n}\h_{\m}+2\f \h_{\m\n}.
\label{transfolin}
\ee
This theory is the linearization of the conformally invariant Weyl gravity 
action 
\be
S^W= \frac{1}{2\l
^2}\int d^4x \sqrt{-g} W_{\a\b\g\d} W^{\a\b\g\d} ~,
\label{actionconf}
\ee
where $ W^{\a}_{~\b\g\d}$ is the conformally invariant Weyl tensor.

As in the previous section, the starting point is really a {\it{sum}} of the
actions (\ref{freeactionconf}), using a quadratic form $k_{ab}$ to perform
the sum.

\subsection{No-go result}

We have the 
\begin{theorem}
\label{t2}
Under the assumptions of : locality, Lorentz invariance, 
a positive sum of actions (\ref{freeactionconf}) as free field limit
and at most four derivatives in the Lagrangian, the only consistent 
deformation of the free action involving a collection of Weyl-gravitons is 
(modulo field redefinitions) a sum of
independent Weyl actions, one for each fields.
\end{theorem}
In this case also, the constant coefficients $a^a_{~bc}$ 
appearing in the deformation were constrained to define a commutative, 
associative, symmetric algebra $\ca$, and together with the
positive-definiteness of the internal  metric $k_{ab}$, it implies
that the interactions between the various Weyl-gravitons can be removed, 
leaving only the self-interactions. 

\subsection{Yes-go result}

In the case of Weyl gravity, there does not appear to be
any particularly strong reason for taking the free
Lagrangian to be a positive sum of free Weyl Lagrangians.
Any other choice, corresponding to an internal metric $k_{ab}$
that need not be definite positive, would seem to be equally good since
the energy is in any case not bounded from below (or above). 
If one allows non positive definite metrics in internal space, then,
non trivial algebras of the type studied in \cite{Cutler:dv}
exist and lead to non trivial cross interactions among the various
types of Weyl gravitons.
We have explicitly given such a two-Weyl-graviton theory in \cite{Boulanger:2001he}.

\section{Non semi-simple extensions of the conformal group}
\label{sect3}

It is well known (see \cite{KTvN1, mansouri}) that the Weyl's theory
of gravity (\ref{actionconf}) is the gauging of the conformal
algebra $so(4,2)$ in four dimensions.
In a forthcoming paper \cite{BHN} we shall show that the two-Weyl-graviton 
theory given in \cite{Boulanger:2001he}
can be expressed as the gauging of the tensor
product $L'=so(4,2)\otimes \ca$, where $\ca$ is the commutative, associative
 algebra encountered in the previous section, with the 
non positive-definite internal metric
$k_{ab}$ making $\ca$ symmetric. 
$\ca$ is spanned by a unit element $e$ and a nilpotent one, $n$, of order two
\footnote{$e^2=e$, $en=ne$, $n^2=0$}. 
Similarly, the gauging of $\tilde{L}=su(2,2\vert 1)\otimes\ca$ 
along the lines of \cite{KTvN2}
gives a superconformal theory with two multiplets in interaction.
One can see that the presence of nilpotent elements in $\ca$ makes the 
tensor product $L'=L\otimes \ca$ no more semi-simple, even if $L$ was 
semi-simple. 

The extension to $N$ multiplets is straightforward : we just
pick out the commutative, associative algebra $\ca$  
spanned by a nilpotent element of 
order $N$, and find the non positive-definite internal metric $k_{ab}$ which 
makes the algebra $\ca$ symmetric.

Although we concentrated on conformal gravity
for the reasons explained above, similar considerations
apply to standard gravity, provided one replaces the conformal
group by the (anti) de Sitter group or its 
Poincar\'e contraction \cite{MacDowell:1977jt}.  In that case
also, one can construct interacting multi-graviton theories
if the metric in internal space is not positive definite
\cite{BDGH,Cutler:dv}. 

Thanks to this formulation in terms of Lie algebras and their gauging,
one observes the following fact : 
all these $N$-field theories can be recovered from the usual
one-field starting theory. Just by doing a Taylor expansion of the fields with
a nilpotent expansion parameter (of order $N$ for a $N$-field theory).
With respect to this observation, it means that even if these multi-field 
theories are allowed
by consistency, they contain no more information than the one-field theory
from which they were built, and thus are in some sense trivial.

\section{Conclusions} 

In this report
we have given results about the consistent deformations of sums of Pauli-Fierz
and linearized Weyl gravity actions, using BRST-cohomology tools 
\cite{Barn-Henn}.
The aim was to build Yang-Mills-like theories of spin-two fields.
Except for an exotic $n=3$ PT-breaking theory, 
such multi-graviton theories are excluded when the sum of free actions 
defining the starting theory is a positive one.
Allowing for non positive sums of free actions as starting point, 
we can build truly interacting multi-graviton theories, but at the light
of a group-theoretical analysis, it becomes clear that these $N$-field
theories are also trivial.

\section*{Acknowledgements} 
The author is very grateful to Marc Henneaux for his 
presence at all stages of the work. The other co-authors cited at the
beginning of the introduction are also acknowledged, together with X. Bekaert
and G. Barnich for kind discussions. 
The work is partially supported by the ``Actions de 
Recherche Concert{\'e}es" of the ``Direction de la Recherche 
Scientifique - Communaut{\'e} Fran{\c c}aise de Belgique", by 
IISN - Belgium (convention 4.4505.86) and by the European Commission 
RTN programme HPRN-CT-2000-00131 in which the author is associated to K. U. Leuven.

\end{document}